\documentclass[prab,twocolumn,showpacs,preprintnumbers,superscriptaddress]{revtex4-2}
\usepackage{graphicx}
\usepackage{amsmath,amssymb}
\usepackage{bm}
\usepackage{hyperref}
\usepackage[utf8]{inputenc}

\begin{document}

\title{The Development of Low-Q Cavity Type Beam Position Monitor with a Position Resolution of Nanometer for Future Colliders}

\author{S. W. Jang}
\email[]{siwon@postech.ac.kr}
\affiliation{Accelerator Department, Pohang Accelerator Laboratory, POSTECH, Pohang 37673, Republic of Korea}

\author{E.-S. Kim}
\affiliation{Accelerator Science Department, Korea University Sejeong Campus, Sejong, Republic of Korea}

\author{T. Tauchi (Retired)}
\author{N. Terunuma}
\affiliation{High Energy Accelerator Research Organization (KEK), Tsukuba, Japan}

\author{P. N. Burrows}
\affiliation{John Adams Institute for Accelerator Science, University of Oxford, United Kingdom}

\author{N. Blaskovic Kraljevic}
\affiliation{Diamond Light Source, Oxfordshire, UK}

\author{P. Bambade}
\author{S. Wallon}
\affiliation{Laboratoire de l'Accélérateur Linéaire (LAL), Orsay, France}

\author{O. Blanco}
\affiliation{Synchrotron SOLEIL, L'Orme des Merisiers, Saint-Aubin, France}

\date{\today}

\begin{abstract}
   The nano-meter beam size in future linear colliders requires very high resolution beam position monitor since higher resolution allows more accurate position measurement in the interaction point. We developed and tested a low-Q C-band beam position monitor with position resolution of nanometer. The C-band BPM was tested for the fast beam feedback system at the interaction point of ATF2 in KEK, in which C-band beam position monitor is called to IPBPM (Interaction Point Beam Position Monitor). The average position resolution of the developed IPBPMs was measured to be 10.1 nm at a nominal beam charge of $87\%$ of ATF2. From the measured beam position resolution, we can expect beam position resolution of around 8.8 nm and 4.4 nm with nominal ATF2 and ILC beam charge conditions, respectively, in which the position resolution is below the vertical beam size in ILC. In this paper, we describe the development of the IPBPM and the beam test results at the nanometer level in beam position resolution.
\end{abstract}

\maketitle

\section{Introduction}
The Accelerator Test Facility 2 (ATF2) at High Energy Accelerator Research Organization (KEK) is a research center for studies on issues concerning the injector, damping ring, and beam delivery system for the ILC~\cite{ILC}.  The  ILC, and ATF2 design parameters are compared in Table~\ref{table1}.
\begin{table}[htb!]
\centering
\caption{\label{table1}ILC TDR and ATF2 parameters.}
\begin{tabular}{lcc}
\hline\hline
Parameter & ILC & ATF2\\ 
\hline
Beam energy (GeV) & 500 & 1.3\\
Number of $e^{-}$ per bunch ($N$) & $2 \times 10^{10}$ & $1 \times 10^{10}$\\
Bunch interval (ns) & 554 & 150 $\sim$ 300\\
Bunch number & 1321 & 60\\
Norm. emittance $\varepsilon_{x}$ (m) & $1 \times 10^{-5}$ &$3 \times 10^{-6}$\\
Norm. emittance $\varepsilon_{y}$ (m) & $3.5 \times 10^{-8}$ &$3 \times 10^{-8}$\\
Beam size $\sigma_{x}$ ($\mu$m) & 0.47 &  2\\
Beam size $\sigma_{y}$ (nm) & 5.9 & 37\\
\hline\hline
\end{tabular}
   \label{table1}
\end{table}

The beam energy of ATF~\cite{ATF} is 1.3 GeV and nominal beam charge is 10$^{10}$ electrons/bunch.
The goal of beam size at the IP region is 37nm vertically, which is the first goal of ATF2. The second goal of ATF2
is to maintain the beam collision with nano meter scale stability at IP-region the vertical beam size in ILC is 5.9nm for the 500GeV. To achieve high beam position resolution of nano meter, we developed prototype low-Q IPBPM and
tested at ATF2 extraction beam line~\cite{proto}~\cite{IEEE}. After the prototype test, we fabricated three low-Q IPBPMs with modified design.
Modified design of the low-Q IPBPM was much smaller and lighter than prototype to install at IP region~\cite{IPBPM}. The entire low-Q IPBPM system consists of three sensor cavities and two reference cavities.
The IPBPM resolution measurement was performed at IP region of ATF2 during beam operation.
Fig.~\ref{fig3} shows the ATF2 layout and the IP area is a location that three IPBPMs and two reference cavity BPMs are installed.
 
 \begin{figure}[!htb]
        \includegraphics[width=0.5\textwidth]{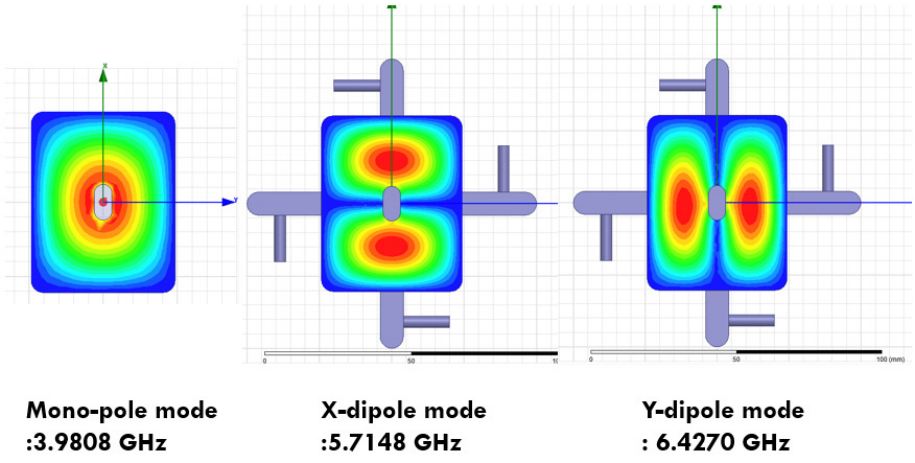}
        \caption{\label{fig3}ATF2 layout, where the ATF2 is the extended test beam line of ATF for the final focus system at future linear collider.}
\end{figure}

\section{Low-Q IPBPM}
 The high resolution low-Q IPBPM will provide the beam position at the IP and will be used for the fast beam feedback system~\cite{fastfeedback} to stabilize the beam orbits of the multi bunches. The low-Q IPBPM was used by rectangular shape cavity to isolate two dipole mode polarizations and the thin cavity reduces the beam angle sensitivity to trajectory inclination~\cite{IEEE}. A position resolution of 8.7 nm was achieved with a high-Q BPM~\cite{highQ} for a beam intensity of 0.7 $\times$ 10$^{10}$ e/bunch with a dynamic range of 5$\mu m$. However, the high-Q IPBPM was not proper to multi-bunch beam operation due to long decay time of RF signal so that we developed a low Q-value cavity BPM to enable the bunch-by-bunch position measurement for the multi-bunch beam with bunch spacing of 154 ns~\cite{proto}. 
 
 \subsection{Development of Low-Q IPBPM}

 As mentioned, we modified design of low-Q IPBPM to install IP region vacuum chamber. The main point of modified design is lighter weight of low-Q IPBPM than prototype BPM. To reduce the weight, the material was changed from copper to aluminium. Moreover, the total cavity size was also reduced from 14cm to 11cm. The dimension of sensor cavity is similar with proto-type low-Q IPBPM but the wave guide part is changed to reduce entire cavity size. The design study of aluminium low-Q IPBPM was performed by using of the electromagnetic simulation program HFSS~\cite{HFSS}. The low-Q IPBPM was used different two dipole modes to avoid isolation issue between x and y port~\cite{highQ}. Therefore, the sensor cavity structure was designed to rectangular shape to split the frequency of two dipole modes.
 
\begin{figure}[!htb]
   \centering
   \includegraphics*[width=72mm]{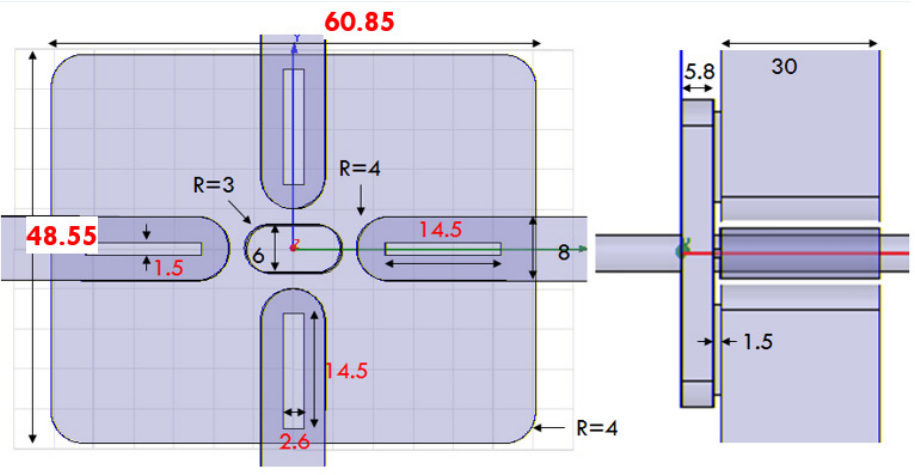}
   \caption{The sensor cavity dimension for HFSS simulation.}
   \label{fig2}
\end{figure}

   The resonant frequencies of two dipole modes are determined by dimension of rectangular cavity in X and Y directions. From the results of electro-magnetic simulation and RF measurements of proto type cavity, the rectangular cavity size was determined. The determined cavity dimension is 60.85mm and 48.55mm for horizontal and vertical direction, respectively. The cavity length L has to be shortened in order to reduce angle sensitivity. However, shorter L decreases R/Q, which reduces position sensitivity also. To recover position sensitivity, a beam pipe radius is required to be small, in order to prevent leakage of the field from the cavity. Finally, the cavity length in the z direction was designed to be 5.8mm and the beam pipe radius are determined to 12mm and 6mm in X and Y direction to achieve the low angle sensitivity and good quality of position sensitivity. Fig.~\ref{fig2} shows the dimension of low-Q IPBPM for HFSS simulation.
 
 The design frequency of two dipole modes of low-Q IPBPM are 5.712 GHz and 6.412GHz for x and y port, respectively. Fig.~\ref{fig3} shows that a monopole field and two dipole mode fields for x and y ports in the sensor cavity. 

\begin{figure}[!htb]
   \centering
   \includegraphics*[width=72mm]{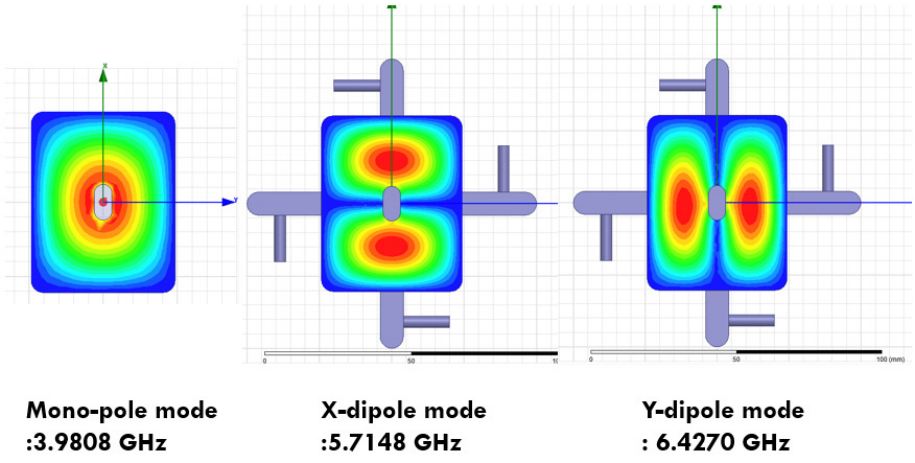}
   \caption{Electric field mapping of HFSS simulation. A resonant frequency of monopole mode is generated at 3.98GHz region(left) and x dipole mode with TM
210 is generated at 5.7148GHz (center) and y dipole mode with TM120  is generated at 6.427GHz (right).}
   \label{fig3}
\end{figure} 

\begin{figure}[htb!]
\includegraphics[width=0.5\textwidth]{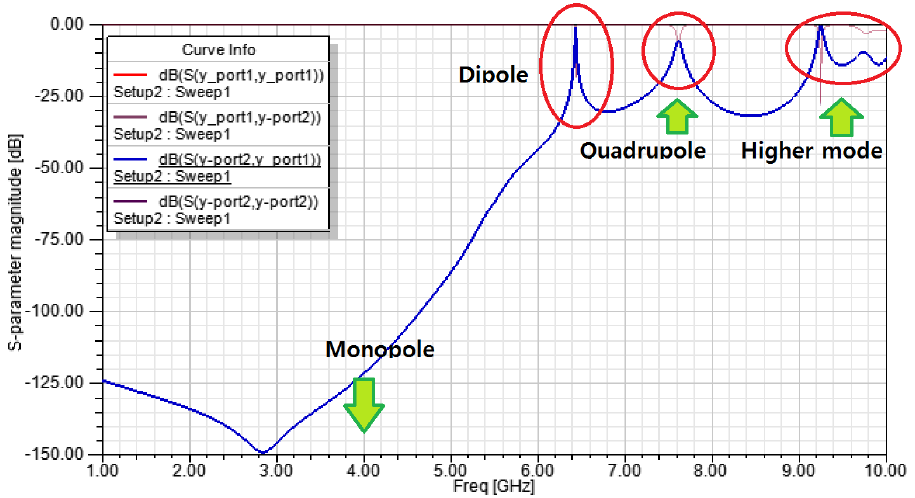}
\caption{\label{fig4}Frequency spectra
of the output signals for the $y$ port by using HFSS~\cite{HFSS}.}
\end{figure}

To get a higher beam position resolution, the dipole mode signal picked up via the feedthrough antenna should be well separate with monopole signal and other higher mode signals. The output signal level of monopole mode at feedthrough antenna is almost negligible as shown in Fig~\ref{fig4}. Because the monopole filed was filtered by cut off frequency of wave guide. The cut off frequency of wave guide was determined by dimension of wave guide and the dimension of modified wave guide is determiend by using follow equation, 

\begin{equation}
f = \frac{c}{2\pi} \times \sqrt{\Big(\frac{m\pi}{a}\Big)^2+\Big(\frac{n\pi}{b}\Big)^2+\Big(\frac{l\pi}{L}\Big)^2}.
\label{eq1}
\end{equation}

By using Eq.~\ref{eq1} and electro-magnetic simulation, the x-port wave guide demension is determined to be $8 \times 30 \times 45 (mm)$ and y-port wave guide demension is determind to be $8 \times 30 \times 42 (mm)$. The other higher modes also well separaeted with dipole field but these higher order mode fields are picked up at feedthrough antenna so that the band pass filter is used to eliminate other higer modes. The isolation of two dipole modes between x-port and y-port was achieved around -50dB as shown in Fig.~\ref{fig4}.

Table~\ref{table2} shows the simulation results of low-Q IPBPM, a resonant frequency of the dipole modes $f_{0}$, the loaded
quality factor $Q_{L}$, the internal quality factor $Q_{0}$, the external quality factor $Q_{ext}$, the coupling constant $\beta$, decay time $\tau$ and transmission parameter $S_{21}$ in dB. The low-Q IPBPM design parameters are calculated by using HFSS simulation~\cite{HFSS}.  

\begin{table}[hbt]
   \centering
   \caption{The design parameters of low-Q IPBPM.}
   \begin{tabular}{lcc}
\hline
\hline
       \textbf{Parameter} & \textbf{x dipole}& \textbf{y dipole} \\
\hline
$f_{0}$ [GHz]     &	5.7148 &	6.4270 \\
$\Delta f$ [MHz]&	7.40    &	11.10\\
$Q_{L}$ 		 &	772 &		579\\
$Q_{0}$ 		 &	4021&	3996\\
$Q_{ext}$ 		 &	956 &		677\\
$\beta$ 		 &	4.2  &	5.9\\
$\tau$ [ns]	 &	21.51  &	14.34\\
$S_{21}$ [dB] 		 &	-1.85 &	-1.36\\
\hline\hline
   \end{tabular}
   \label{table2}
\end{table}

\subsection{Design of the reference cavity BPM}

\begin{figure}[!htb]
   \centering
   \includegraphics*[width=80mm]{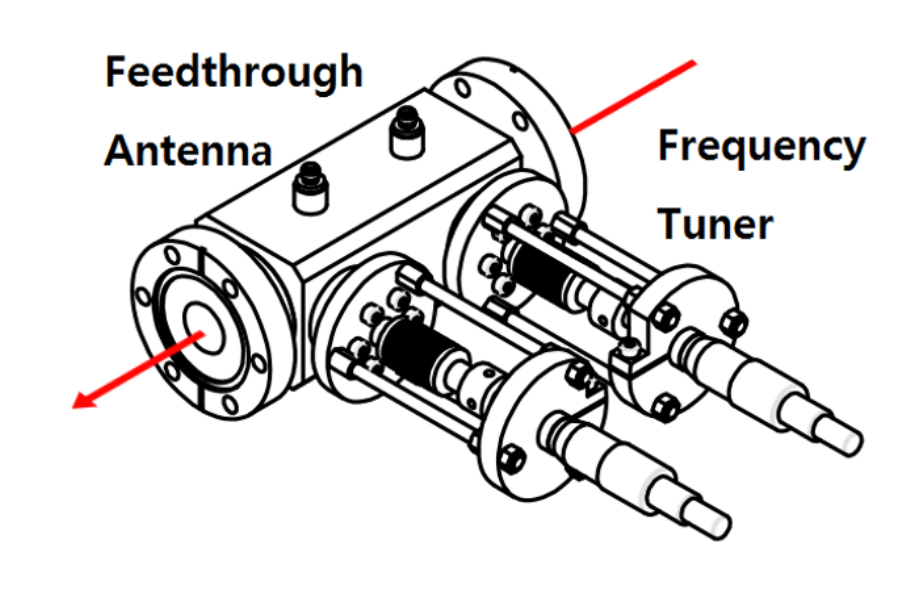}
   \caption{The design of reference cavity BPM for low-Q IPBPM.}
   \label{fig5}
\end{figure}

To obtain the beam phase reference and measure the beam charge, the cylinderical shape reference cavity BPMs were designed with the
monopole mode, TM010, and the same frequencies as the two dipole modes of the low-Q  IPBPM. Since excitation of
the monopole mode dominates all other modes so that there is no special selective coupler. As shown in Fig.~\ref{fig5}, the excited
signal of the reference cavity is coupled out to a feedthrough antenna through a small wave guide. The two reference cavities are 
designed to correspond to resonant frequencies of the low-Q IPBPM's two dipole modes. To matched resonant freqencies of the low-Q IPBPM, the diameter
 of reference cavities are determined to be 42.95mm and 38.65mm for x-port and y-port, respectively.
 
\begin{figure}[!htb]
   \centering
   \includegraphics*[width=72mm]{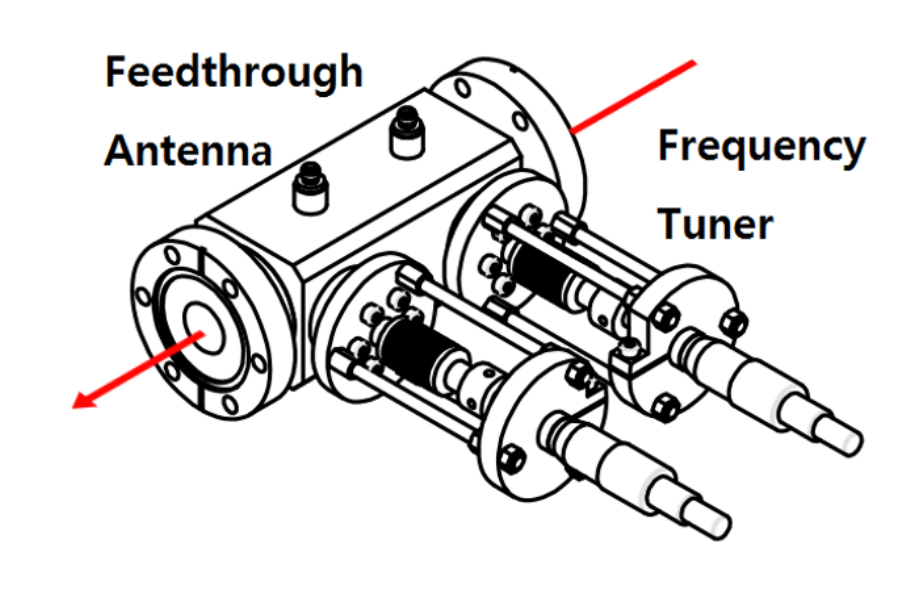}
   \caption{The technical drawings of reference BPM.}
   \label{fig6}
\end{figure}

The reference cavity BPM has frequency tuner for the frequency matching with average of three low-Q IPBPMs. In theory, the resonant frequencies of the manufactured  three beam position monitors and the reference cavity BPM should be the same as the results of the HFSS simulations, but the actual resonant frequencies of each BPM are different. Therefore, the reference cavity BPM should be matched to average resonant frequency of three cavity BPMs by using frequency tuner. Fig.~\ref{fig6} shows the technical drawing of reference cavity BPM. The reference cavity BPM consists of two reference cavities and two frequency tuner with $\pm$3MHz range.

\subsection{Fabrication of the Low-Q IPBPM}

\begin{figure}[!htb]
   \centering
   \includegraphics*[width=72mm]{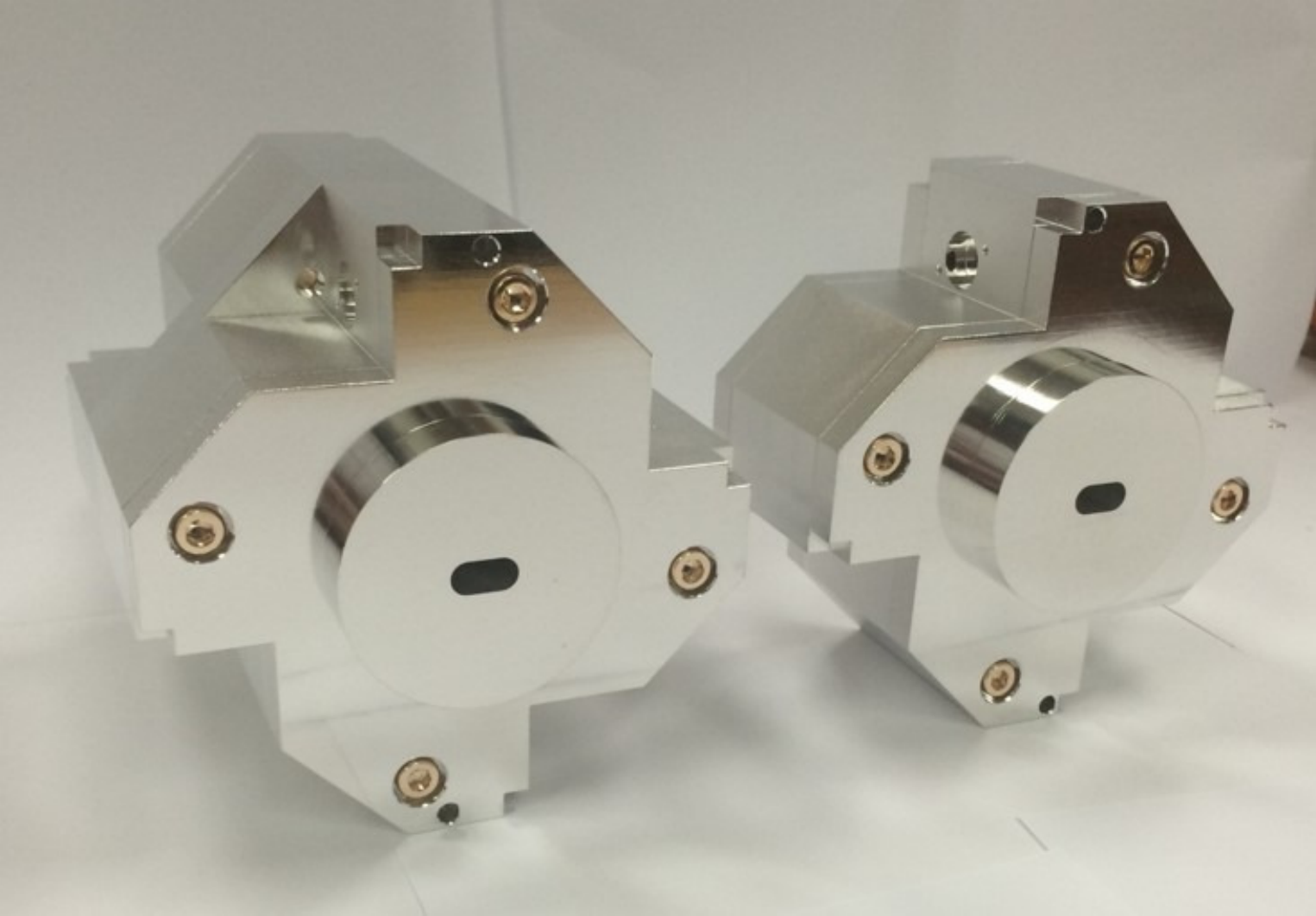}
   \caption{The fabricated low-Q IPBPM. There are two block of IPBPMs. Left one shows a double IPBPM block and right one shows a single IPBPM block.}
   \label{fig7}
\end{figure}

The install version of low-Q IPBPM was fabricated and tested. At the first time of fabrication, we found that the chemical polishing for the clean surface of low-Q IPBPM
cavity made irregular surface flatness and which leads to performance degration. For the second fabrication of low-Q IPBPM we do not performed chemical polishing and additionally
performed indium sealing on the cavity surface to avoid leakage of RF field between cavity part and cavity cover. Figure~\ref{fig7} shows the fabricated low-Q IPBPM blocks.

After the second fabrication and indium seal, the $Q_{L}$ of low-Q IPBPM was increased so that the decay time was recovered to enough for the digitization of analog signal.
Table~\ref{table3} and ~\ref{table4} show the RF test results of first fabrication, before indium sealing of second fabrication and after indium sealing case of IPBPM-C. 
The Y-port parameters are improved well but X-port parameters are improved slightly. However, the decay time has been recovered enough to be digitized.

\begin{table}[hbt]
   \centering
   \caption{The RF test results of low-Q IPBPM.}
   \begin{tabular}{lccc}
\hline
\hline
       \textbf{Y-port} & \textbf{1st fab}& \textbf{2nd fab} & \textbf{2nd fab} \\
       \textbf{Parameter} & \textbf{w/ CP}& \textbf{w/o CP} & \textbf{w/ indium seal} \\
\hline
$f_{0}$ [GHz]     &	6.4165 &	6.4255 &	6.42175 \\
$\Delta f$ [MHz]&	24.0   &	11.52 &	8.22\\
$Q_{L}$ 		 &	267 &		557 &		781\\
$Q_{0}$ 		 &	334&		1107 &	2311\\
$Q_{ext}$ 		 &	1186 &	1123 &	1180\\
$\beta$ 		 &	0.29  &	0.986 &	1.959\\
$\tau$ [ns]	 &	6.623  &	13.816 &	19.362\\
S21 [dB] 		 &	-13.15 &	-6.083  &	-3.583\\
\hline
   \end{tabular}
   \label{table3}
\end{table}

\begin{table}[hbt]
   \centering
   \caption{The RF test results of low-Q IPBPM.}
   \begin{tabular}{lccc}
\hline
\hline
       \textbf{X-port} & \textbf{1st fab}& \textbf{2nd fab} & \textbf{2nd fab} \\
       \textbf{Parameter} & \textbf{w/ CP}& \textbf{w/o CP} & \textbf{w/ indium seal} \\
\hline
$f_{0}$ [GHz]     &	5.7145 &	5.7131 &	5.712 \\
$\Delta f$ [MHz]&	29.0   &	15.42 &	12.5\\
$Q_{L}$ 		 &	197 &		371 &		457\\
$Q_{0}$ 		 &	453&		1202 &	2102\\
$Q_{ext}$ 		 &	348 &		535 &		584\\
$\beta$ 		 &	1.17  &	2.283 &	3.598\\
$\tau$ [ns]	 &	5.487  &	10.34 &	12.733\\
S21 [dB] 		 &	-5.35 &	-3.171  &	-2.135\\
\hline
   \end{tabular}
   \label{table4}
\end{table}

\section{Installation of Low-Q IPBPM with piezo mover}

\begin{figure}[!htb]
   \centering
   \includegraphics*[width=72mm]{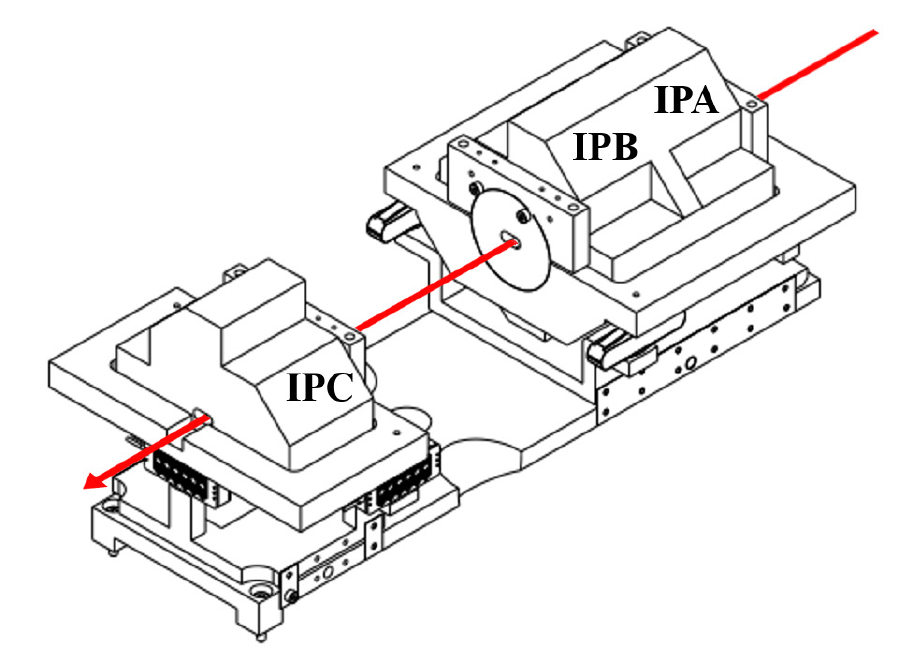}
   \caption{The installed three low-Q IPBPMs on the base
plate. There are two different piezo mover systems to control
BPM position.}
   \label{fig8}
\end{figure}

Three low-Q IPBPMs were installed inside the IP-chamber. To install inside IP-chamber, three low-Q IPBPM cavities are
first installed on the base plate as shown in Fig.~\ref{fig8}, which base plate with piezo mover system was developed by LAL group.
Two different type of piezo mover system are installed bottom of two cradle for the IPBPM block support to control BPM
position in horizontal and vertical directions. The precise position alignment of low-Q IPBPM block was performed
by using these piezo mover system. We used two different piezo mover type, one is manufactured by Cedrat and the
other one is manufactured by PI. A Cedrat piezo movers support a double IPBPM block with $250\mu m$ operation
range and PI piezo movers support a signle IPBPM block with $300\mu m$ operation range. Two types of piezo-mover
were set up four each. The one of piezo mover is installed in lateral direction to control horizontal direction of low-Q
IPBPM and the other three piezon movers are installed in vertical direction to control vertical direction and rotation,
the rotation of IPBPM block can be adjusted by combination of three different vertical piezo positions.

\section{SIGNAL PROCESSING OF LOW-Q IPBPM WITH ELECTRONICS}

\begin{figure}[!htb]
   \centering
   \includegraphics*[width=80mm]{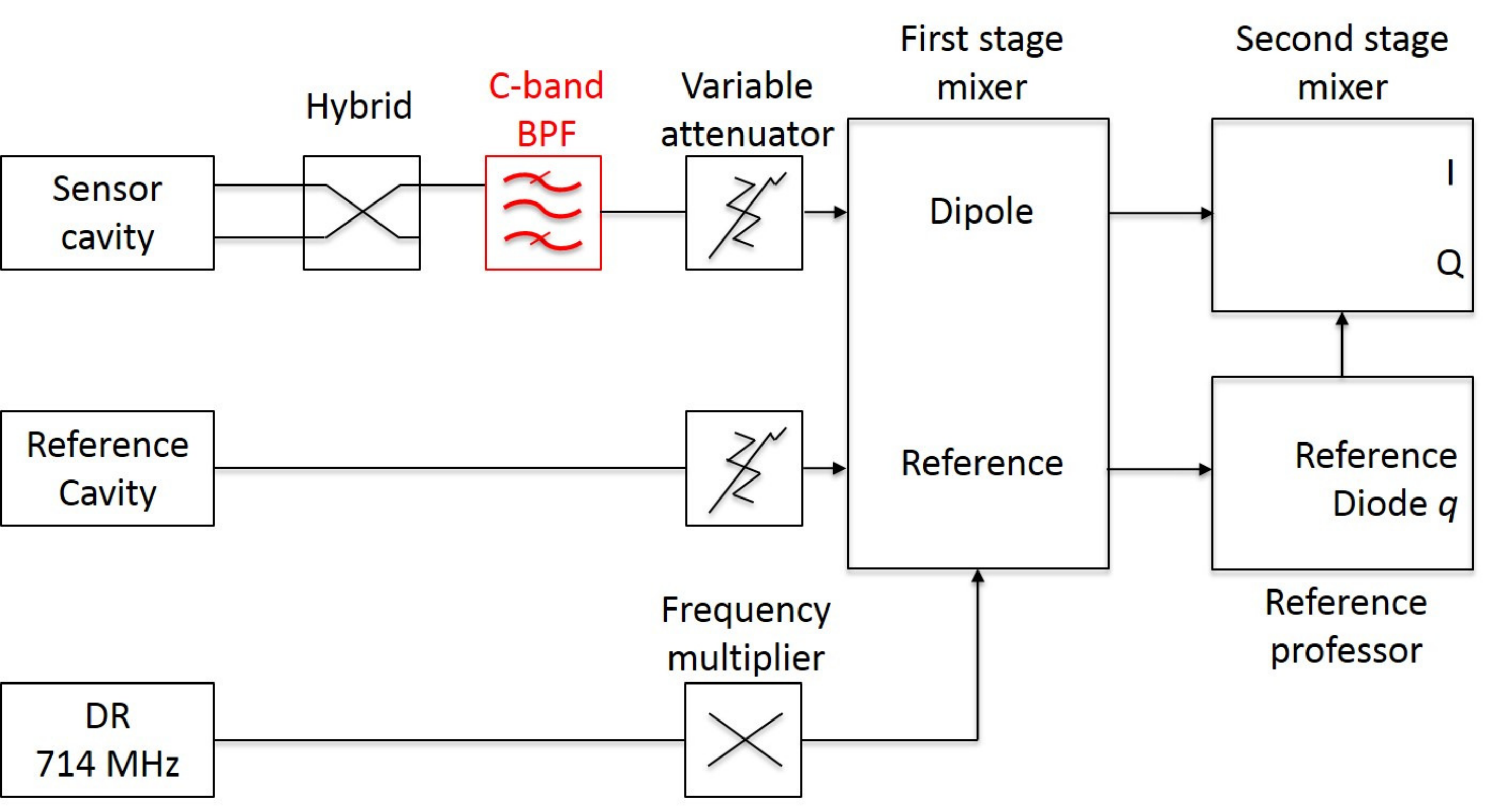}
   \caption{The beam test scheme for low-Q IPBPM entire system.}
   \label{fig9}
\end{figure}

Three low-Q IPBPMs, two reference cavity BPMs, 1st stage electronics, LO signal splitter, C-band BPF, hybrid
and variable attenuators( x8) are installed inside tunner near the IP-region of ATF. At the outside of tunnel, the 2nd stage
electronics and ADC are installed. More detailed scheme was shown in the Fig.~\ref{fig9}. First, an excited RF signal from
low-Q IPBPM cavity pass through the hybrid to combined 180 degree phase diffence two signals in each horizontal
and vertical direction and then passes through C-band band pass filter to eliminated remaind mono-pole signal and other
higher mode signals. A filtered RF signal amplitude can be attenuated by using variable attenuator from 0dB to 70dB
with 10dB step, which attenuator will be used for the wide dynamic range beam position calibration but will not used for
the beam position resolution measurement. The RF signal into the first stage electronics [7] for the signal amplification
and signal mixing with multiplied local oscillator(LO) signal, which 714 MHz LO signal is provided by the ATF damping
ring rf system and it is used to generate the LO of 5712 MHz by using 8th harmonics generator. After the first stage of
electronics the RF signal frequency is changed to 714MHz of IF signal to avoid signal power loss during 30m long cable
transfer section from inside tunnel to outside tunnel.

The signal processeing of reference cavity in the first stage electronics is similar with RF signal. A BPF is used
to reject the other modes and then the signal is mixed with same external LO signal, it also is down-converted to 714
MHz while conserving the phase relation with the IPBPM RF signals. The IF signals from first stage electronics are
further processed using second stage electronics, which is IQ phase detector module [7]. In the IQ phase detector module,
the IF signal is split and sent to two mixers and then detected into the base band having orthogonal phases, which is I and
Q signals with 90 degree differences. By using these I and Q signal we can measure beam position and angle informations.
The bandwidth is determined by a 100 MHz low-pass filter placed after the mixer. The phase origin of the detection can
be adjusted using the manual phase shifter at the input of the phase reference.

\section{THE PRINCIPLE OF POSITION RESOLUTION MEASUREMENT}

\begin{figure}[!htb]
   \centering
   \includegraphics*[width=80mm]{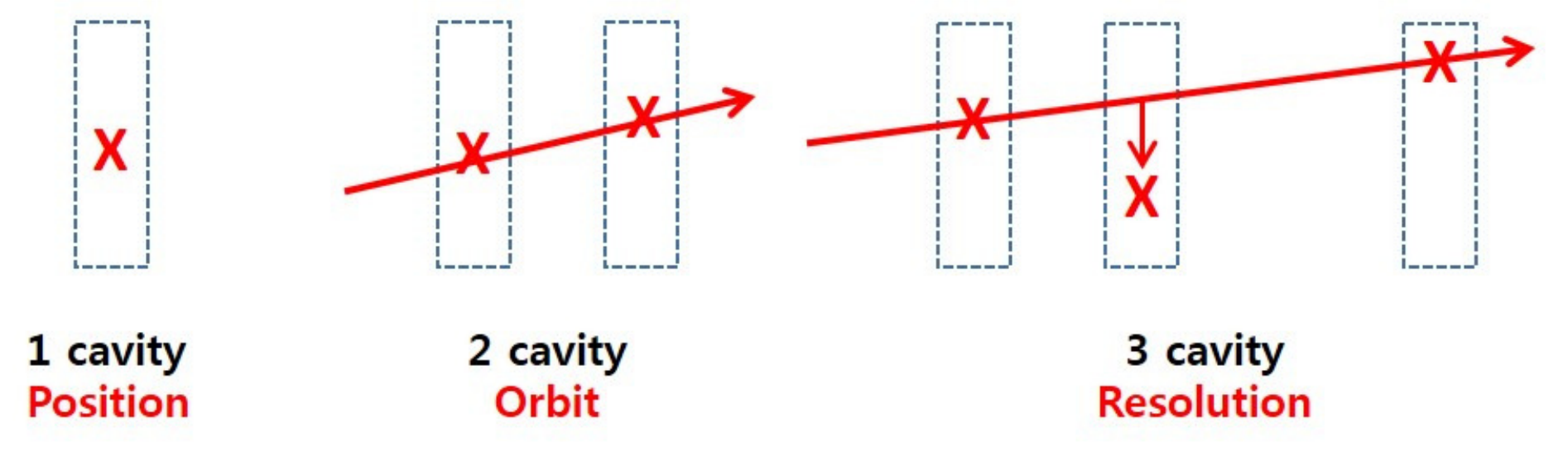}
   \caption{The principle of beam position resolution measurement of IPBPM.}
   \label{fig10}
\end{figure}

  As shown in Figure~\ref{fig10}, single BPM can be determined a beam position and two BPMs can be determined beam orbit. We can measure the beam position resolution by using three cavity BPMs. 
 Therefore, three low-Q IPBPMs are used for the measurement of the beam position resolution. First, two BPMs are used to find the predicted position by calculating the beam orbit and then we can calculated the RMS of residual between the measured beam position and calculated predict beam position. Finally, the beam position resolution was determined by ``The RMS value of the residual position at the low-$Q$ cavity BPM" $\times$ ``geometrical 
factor." The geometrical factor was used to correct for propagation of the error. Also, we assumed that the three cavities had the same position resolution.

 The position resolution measurement consists of the following 3 steps,
 
\begin{itemize}
	\item  I-Q phase tuning, to distinguish the position signal as I and the other noise component as Q.
	\item  Calibration Run, to calibrate the sensor cavity signal amplitude due to different beam position level.
	\item  Resolution Run, to measure the RMS of the residual between measured and predicted beam position at low-Q IPBPM.
\end{itemize}

\section{CALIBRATION RUN OF LOW-Q IPBPM}

\begin{figure}[!htb]
   \centering
   \includegraphics*[width=70mm]{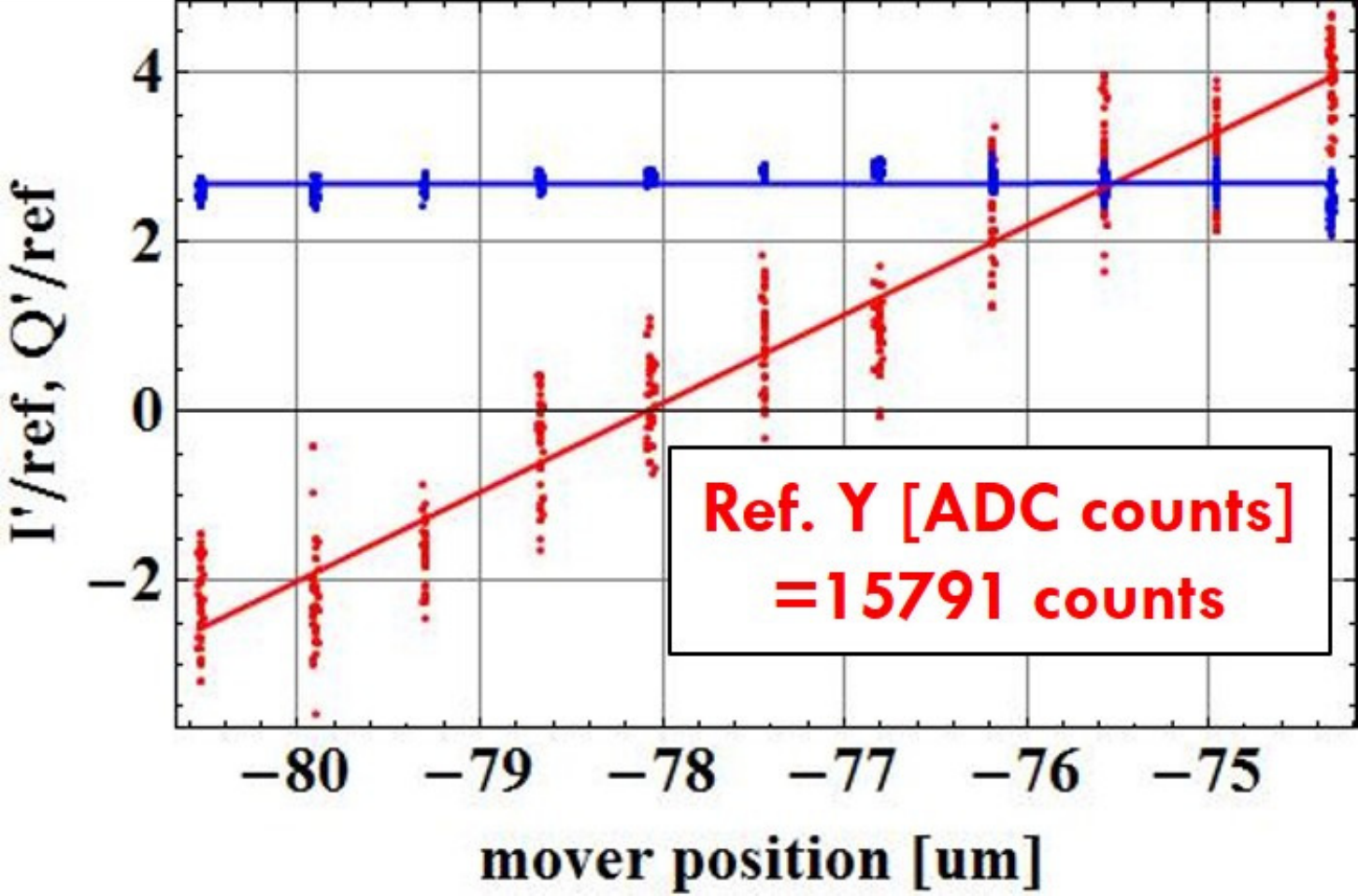}
   \caption{low-Q IPBPM-A calibration factor. Normalized I' signal (Red) and normalized Q' signal(Blue).}
   \label{fig11}
\end{figure}

The calibration run was performed to calibrate the sensor cavity response to actual beam position. 
The sensor cavities of low-Q IPBPM are swept against the electron beam orbit by controlling piezo mover system and response of the output voltage of sensor cavities are monitored. Two type of piezo mover system are installed at IP region and the dynamic range of piezo mover system are $300\mu m$ for PI and $250\mu m$ for Cedrat with nano meter level accuracy. The calibration run took 20 data at each mover position. To calculate the calibration factor, we first calculate the normalized I' signal and Q' signal. The normalized I' signal and Q' signal are determined by the following step. 
\begin{itemize}
    \item   $I' = (I \times Cos\theta + Q\times Sin\theta)/(Ref. signal),$
    \item  $Q' = (Q \times Cos\theta - I\times Sin\theta)/(Ref. signal),$
\end{itemize}
where the $\theta$ means the IQ rotation angle. Even though we performed IQ phase tuning to distinguish the position signal and noise component, the Q signal  still include small amount of the position information. Therefore, we should calculate the IQ rotation angle to calculate actual position signal and noise component. 
The calibration factor was calculated by using integration method from sample number $\#53$ to $\#59$. The Fig.~\ref{fig11} shows the results of calibration run for low-Q IPBPM A case. The low-Q IPBPM calibration factors are listed in Table\ref{table5}.

\begin{table}[hbt]
   \centering
   \caption{The calbration factor of low-Q IPBPM.}
   \begin{tabular}{lcc}
\hline
\hline
       \textbf{Channel} & \textbf{Cal factor}& \textbf{Norm. cal factor} \\
       \textbf{of BPM} & \textbf{[ADC counts/um]}& \textbf{[/um]} \\
\hline
IPA YI     &	16599 &	1.0512 \\
IPB YI     &	12062 &	0.7639 \\
IPC YI     &	7810 &      0.4946\\
\hline
\hline
   \end{tabular}
   \label{table5}
\end{table}

\section{Resolution run of low-Q IPBPM}
The position resolution of  the low-$Q$ cavity BPM was estimated with a fixed beam offset. 
The electronics setup was used same configuration as in the calibration run. The main purpose of resolution run was to measure the residual, which is the difference between the measured position at one of BPM and the predicted position by using the other two BPMs. 
The predicted beam position was obtained from a linear regression analysis by using informations from IPBPM-B(IPB) and IPBPM-C(IPC). Those 10 parameters are,
\begin{itemize}
\item   Vertical position signals (in phase components):\\ IPB-YI, IPC-YI
\item   Vertical noise components (out of phase components):\\ IPB-YQ, IPC-YQ
\item   Horizontal position signals and noise components:\\ IPB-XI, IPC-XI, IPB-XQ, IPC-XQ
\item   Beam charge detected at X $\&$ Y reference cavity:\\ Ref-X, Ref-Y
\end{itemize}

The linear regression formula by using 11 parameters was shown in below, and determined the coefficient $\alpha$ of each parameter.
\begin{itemize}
\item  $IPAYI^{\prime} = \alpha0+\alpha1\cdot IPBYI^{\prime}+ \alpha2\cdot IPBYQ^{\prime}+\alpha3\cdot IPCYI^{\prime}$+ $\alpha4\cdot IPCYQ^{\prime}+ \alpha5\cdot RefY+\alpha6\cdot IPBXI^{\prime}+\alpha7\cdot IPBXQ^{\prime}$+$\alpha8\cdot IPCXI^{\prime}+ \alpha9\cdot IPCXQ^{\prime}+ \alpha10\cdot RefX$
\end{itemize}

The residual value can be calculated as follows: 
\begin{eqnarray}
Residual =  Y_{I_{meas}} - Y_{I_{predicted}}.
\end{eqnarray}

\begin{figure}[!htb]
   \centering
   \includegraphics*[width=85mm]{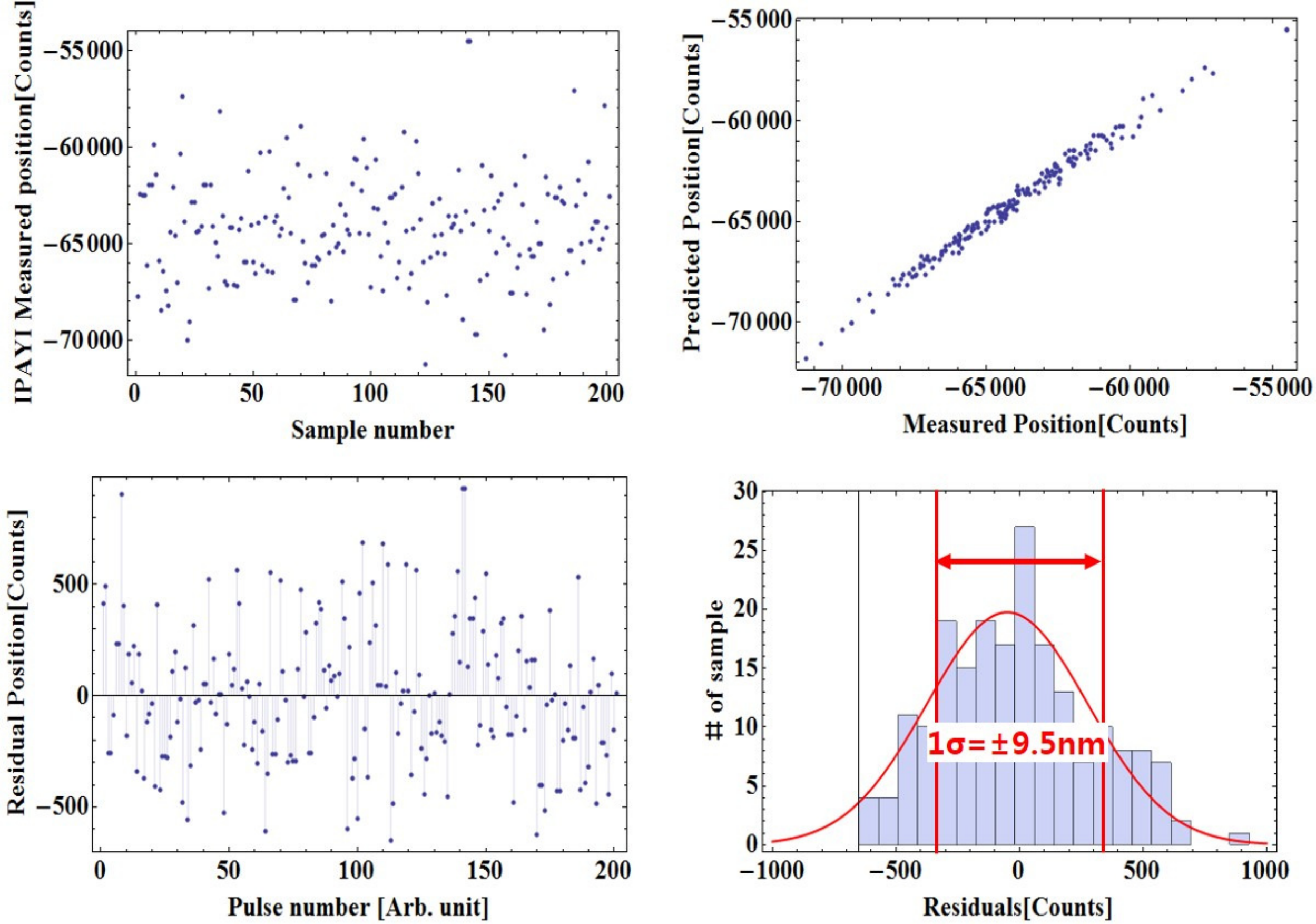}
   \caption{The measured position resolution of IPBPM-A is 9.5nm in beam charge condition of 0.869 $\times$ 1.6nC. }
   \label{fig12}
\end{figure}

Figure~\ref{fig12} shows the result of the resolution run under 0 dB attenuation. Left top of Fig.~\ref{fig12} shows the measured beam position at low-Q IPBPM-A and right top of Fig.~\ref{fig12} shows the measured position vs predicted position of IPBPM-A. The calculated residual(left bottom) and the distribution of residual(right bottom) are shown in bottom of Fig.~\ref{fig12}. The RMS of the residual, 330.638 ADC counts, corresponds to the position resolution.

By using the RMS of residual, we can calculate the beam position resolution as follows:
\begin{equation}
\mathrm{fig12} = \mathrm{Geo. factor}\times \frac{\mathrm{RMS\,\, of\,\, residual}}{\mathrm{calibration\,\,\, factor}}.
\label{eq1111}
\end{equation}

\begin{figure}[!htb]
   \centering
   \includegraphics*[width=85mm]{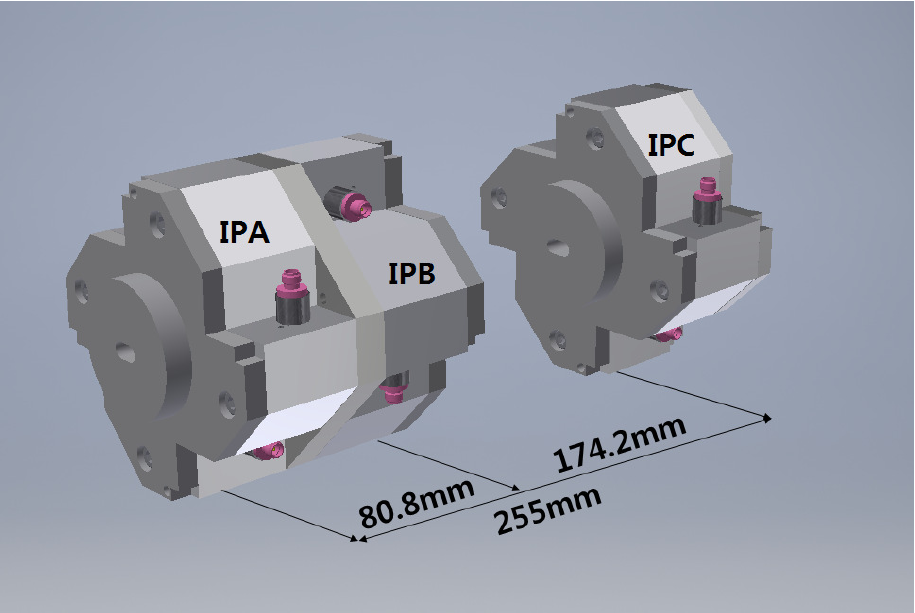}
   \caption{The distances between each low-Q IPBPMs at IP-area. The distances are measured from one of the BPM cavity center to other BPM cavity center.}
   \label{f7}
\end{figure}

The extrapolating method by using geometrical relation between three IPBPMs was used. Fig.~\ref{f7} shows the distance between each BPM.

The distance between BPM: $Z_{12}$ = 80.8mm, 
$Z_{23}$ = 174.2mm, $Z_{13}$ = 255mm. Also, we define $BPM_{i}$ as the i th cavity, $Z_{ij}$ as the distance between $BPM_{i}$ and $BPM_{j}$, $I_{i}$ as beam position signal $BPM_{i}$ and assuming their resolution $R_{i}$ are all equivalent, RMS of

\begin{displaymath}
Z_{12} : (I_{2}- I_{1}) = Z_{13} : (I_{3}-I_{1})
\end{displaymath}
\begin{displaymath}
\Rightarrow	Z_{13}(I_{2}- I_{1}) = Z_{12}(I_{3}-I_{1})\\
\end{displaymath}
\begin{displaymath}
\Rightarrow	Z_{13}I_{2} = Z_{12}I_{3} + (Z_{13} + Z_{12})I_{1}       \\
\end{displaymath}
\begin{displaymath}
\Rightarrow	Z_{13}I_{2} = Z_{12}I_{3} + Z_{23}I_{1}     \\
\end{displaymath}
\begin{displaymath}
\Rightarrow	I_{2} = (Z_{12}I_{3} + Z_{23}I_{1})/ Z_{13}    \\ 
\end{displaymath}
\begin{displaymath}
\therefore f_{B} ( I_{1}, I_{2}, I_{3}) =  I_{2} - (Z_{12}I_{3} + Z_{23}I_{1})/ Z_{13}\quad	\textrm{for IPBPM-B}\\
\end{displaymath}
\begin{displaymath}
 \quad f_{A}( I_{1}, I_{2}, I_{3}) = I_{1} - (Z_{13}I_{2} -Z_{12}I_{3})/ Z_{23}       \quad\textrm{for IPBPM-A}\\
\end{displaymath}
\begin{displaymath}
\quad f_{C} ( I_{1}, I_{2}, I_{3}) = I_{3} - (Z_{13}I_{2} - Z_{23}I_{1})/ Z_{12}       \quad\textrm{for IPBPM-C}\\
\end{displaymath}

The first term of function is the measured $I_{i}$ value and the second term is the predicted $I_{i}$ value, in which predicted value is interpolated by $I_{j}$ and $I_{k}$ .
 Then the Geometrical factor of IPBPM-B is calculated as \cite{nakamura},

\begin{equation}
\frac{R_2}{R_f}=R_2\Big/\sqrt{\left(\frac{\partial f}{\partial I_1}R_1\right)^2+\left(\frac{\partial f}{\partial I_2}R_2\right)^2+\left(\frac{\partial f}{\partial I_3}R_3\right)^2}
\end{equation}
\begin{equation}
=1\Big/\sqrt{\left(\frac{Z_{23}}{Z_{13}}\frac{R_1}{R_2}\right)^2+1+\left(\frac{Z_{12}}{Z_{13}}\frac{R_3}{R_2}\right)^2}
\end{equation}
\begin{equation}
=1\Big/\sqrt{\left(\frac{Z_{23}}{Z_{13}}\right)^2+\left(\frac{Z_{12}}{Z_{13}}\right)^2+1}
\end{equation}

In theoretically, the beam position resolution of three low-Q IPBPM is equal due to used same design so that we assumed $R_{1}$ = $R_{2}$ = $R_{3}$.  Therefore, the Geometrical factor for IPBPM-B is approximately 0.7988. 
 The used geometrical factors for each BPM are shown in the Table~\ref{table6}. 
\begin{table}[hbt]
   \centering
   \caption{The geometrical factor of low-Q IPBPM.}
   \begin{tabular}{l ccc }
\hline
\hline
       \textbf{} & \textbf{IPBPM-A}& \textbf{IPBPM-B} & \textbf{IPBPM-C}\\
\hline
Geo. factor&	0.5457&	0.7988&	0.2531\\ 
\hline
\hline
   \end{tabular}
   \label{table6}
\end{table}

The results of beam position resolution measurement of low-Q IPBPM are summarized in Table~\ref{table7}. The measured average position resolution was 10.1nm with 0.87$\times$ 10$^{10}$ e/bunch. This measured position resolution implies that the normalized beam position resolution with nominal beam condition of ATF, which is 1.00$\times$ 10$^{10}$ e/bunch, is expected to be 8.8nm.

\begin{table}[hbt]
   \centering
   \caption{The measured and expected resolution of low-Q IPBPM.}
   \begin{tabular}{l ccc }
\hline
\hline
       \textbf{} & \textbf{IPBPM-A}& \textbf{IPBPM-B} & \textbf{IPBPM-C}\\ 
\hline
Meas. resol. &	9.50nm&	11.2nm&	9.77nm\\ 
Norm. resol. &	8.26nm&	9.77nm&	8.50nm\\ 
\hline
\hline
   \end{tabular}
   \label{table7}
\end{table}

\section{Conclusion}
In this paper, we described the development and the results of beam test of a low-$Q$ IPBPM. The low-Q IPBPM was developed to provide the beam position information at the IP and will be used for the fast beam feedback system to stabilize the beam orbits of the multi bunches for linear colliders. The measured average beam position resolution was 10.1nm for 0.87$\times 10^{10}$ e/bunch and the expected resolution for nominal beam charge of $2\times10^{10}$ e/bunch in ILC case was 4.4nm in the vertical direction.

\end{document}